\documentclass{ifacconf}

\usepackage{graphicx}      
\usepackage{natbib}        
\usepackage{amssymb}
\usepackage{amsmath}

\usepackage{graphicx}      
\usepackage{natbib}        
\usepackage{algorithm,algorithmicx}
\usepackage[noend]{algpseudocode}
\newtheorem{remark}{Remark}

\usepackage{xcolor,color}
\definecolor{blu}{rgb}{0,0,1}

\definecolor{gre}{rgb}{0,.5,0}

\definecolor{red}{rgb}{1,0,0}

\newcommand{\email}[1]{{\small\tt #1}}
\newcommand{\E}{{\mathbb E}}
\begin{document}
\begin{frontmatter}

\title{Reinforcement Learning with Partial Parametric Model Knowledge} 
\thanks[footnoteinfo]{\textcopyright 2023 the authors. This work has been accepted to IFAC World Congress for publication under a Creative Commons Licence CC-BY-NC-ND. The authors gratefully acknowledge the financial support of Natural Sciences and Engineering Research Council of Canada and Honeywell Process Solutions.}


\author[First]{Shuyuan Wang} 
\author[Second]{Philip D. Loewen} 
\author[Second]{Nathan P. Lawrence}
\author[Third]{Michael G. Forbes}
\author[First]{R. Bhushan Gopaluni}

\address[First]{
Department of Chemical and Biological Engineering, 
University of British Columbia, Vancouver, BC V6T 1Z3, 
Canada \\
(e-mail: 
\email{antergravity@gmail.com},
\email{bhushan.gopaluni@ubc.ca}
)}
\address[Second]{
Department of Mathematics, 
University of British Columbia, Vancouver, BC V6T 1Z2, Canada \\
(e-mail: 
\email{loew@math.ubc.ca},
\email{input@nplawrence.com}
)}
\address[Third]{
Honeywell Process Solutions, North Vancouver, BC V7J 3S4,
Canada \\
(e-mail: 
\email{michael.forbes@honeywell.com}
)}

\begin{abstract}                
We adapt reinforcement learning (RL) methods for continuous control to bridge the gap between complete ignorance and perfect knowledge of the environment. Our method, Partial Knowledge Least Squares Policy Iteration (PLSPI), takes inspiration from both model-free RL and model-based control. It uses incomplete information from a partial model and retains RL's data-driven adaption towards optimal performance. The linear quadratic regulator provides a case study; numerical experiments demonstrate the effectiveness and resulting benefits of the proposed method.
\end{abstract}


\end{frontmatter}

\section{Introduction}
Reinforcement learning (RL) is an online process that enables an agent to autonomously adapt to its environment. 
It has recently achieved great success on complex tasks such as playing Atari games \citep{atari}, winning at Go \citep{go}, and OpenAI Five \citep{openai5}. 
However, RL is famously data-hungry \citep{mastering_atari,samplerlsurvey}. 
Poor sample efficiency is a secondary concern in the purely virtual tasks mentioned above, but it is a critical limitation in real-world systems like robotics, healthcare, and industrial processes, where every observation costs time and money.

Researchers have made a lot of progress in sample efficient RL in recent years, including exploration \citep{exploration}, environment modeling \citep{modeling}, abstraction \citep{abstraction}, meta-RL \citep{mcclement2022meta} and more. In most of these cases, however, the RL formulation assumes the learner knows nothing of the system. In many practical problems, we have some prior information about the system, such as its order, structure, or even some of the model parameters. If RL could utilize such information instead of learning everything about its environment from scratch, we can reasonably expect an improvement in sample efficiency.

Of course, when a complete and accurate model for the system is available, classic model-based control techniques, such as optimal control, will be preferred. Sampling and data are not required at all. One even has tidy analytical solutions in certain well-studied cases, such as the linear quadratic regulator (LQR). 
These classic control methods do not adapt to changes in the environment, however. Furthermore, complete and accurate models are vanishingly rare in practice \citep{spielberg2019toward}. In this research, we aim to build a bridge from the theoretical world of LQR to the practical world of partial knowledge and inevitable uncertainty.




The recent advances in model-based RL (MBRL) share the motivations above \citep{pilco}. 
MBRL was also developed to improve sample efficiency,
however, sample efficiency is achieved by trading off performance.
In spite of the term ``model-based" in its name, 
MBRL works in a different way from control.
It first uses sampled data to fit a model, then uses the synthetic model to generate more data. These artificial data and real data are merged into a single batch of synthetic data to train a policy. Hence, the fundamental idea behind MBRL is still feeding the learning agent with more data. In contrast, this research aims to develop a learning framework using \textit{sample-free} control techniques while retaining the ability to explore the uncertain part of the environment in pursuit of optimal performance.

Within the limited research on this topic, \citep{tamar2011integrating,pkmdp} define and use partial model knowledge in an RL framework, but their methods are based on discrete Markov decision process (MDP) models with finite action and state spaces. This does not fit with continuous control. 

In the initial explorations described here, we focus on LQR, but we expect nonlinear applications to follow. In practice, LQR models are often used to approximate and control more general systems. Further, even though LQR is the most basic and important optimal control problem with unbounded, continuous state and action spaces, the problem of incorporating \emph{a priori} information has not been thoroughly investigated.

To this end, we propose a novel off-policy algorithm named Partial Knowledge Least Squares Policy Iteration (PLSPI) for learning-based LQR with partially known parametric model information. Specifically, our method combines classic LSPI \citep{LSPI}, which is a model-free method utilizing a linear structure of the value function, with optimal control techniques.

The method proposed here offers two key benefits:

\begin{enumerate}
    \item PLSPI gives RL the ability to use partial parametric system information, so that the sample efficiency can be improved.
    \item PLSPI provides a conceptual link between model-free learning and model-based control.
\end{enumerate}

\section{Preliminaries}\label{sec.II}
\subsection{Notation}
For any fixed vector $q$ in $\mathbb{R}^n$, 
the quadratic form $q^T Q q$ is a linear function
of its symmetric $n\times n$ matrix $Q$.
To make this explicit, 
define the ``overbar'' function on vectors and matrices.
Let $\overline{q}$ denote the column vector of length 
$\frac12n(n+1)$ whose elements list all possible
products of two elements from $q$, i.e., \[\overline{q}=[q_1^2,q_1q_2,\dots,q_1q_n,q_2^2,q_2q_3,\dots,q_2q_n,q_3^2,\dots,q_n^2]^T.
\]
Then let $\overline{Q}$ denote the vector of length  $\frac12n(n+1)$ whose elements list the upper triangular entries in $Q$, ordering the elements of $\overline{q}$ and $\overline{Q}$ so that the usual quadratic form involving $q$ and $Q$ turns into an inner product:
\[
q^T Q q =\overline{q}'\overline{Q}.
\]
Use $\mathbf{Tr(\cdot)}$ to represent the trace of a matrix; and $\E(\cdot)$ for the mathematical expectation; let $[\cdot]^T$ and $[\cdot]'$ equivalently represent transpose of a vector or a matrix.

\subsection{The linear quadratic regulator (LQR)}
LQR aims to minimize the accumulated quadratic cost for a linear dynamical system. This paper focuses on the infinite horizon time-invariant discrete-time LQR, in which the cost function is

\[
r(x_t,u_t)=x_t^TQx_t+u_t^TRu_t
\]
for prescribed matrices $Q=Q^T\geq0$ and $R=R^T>0$.
In the general formulation, a discount factor $\gamma\in(0,1]$ is given and the problem is
\begin{equation}
	\begin{aligned}
		&\text{min}\quad   &J= \E\left(\sum_{t=0}^{\infty} \gamma^t r(x_t,u_t)\right)\\
		&\text{s.t.} \quad &x_{t+1}=Ax_t+Bu_t+\xi_t
	\end{aligned}
\end{equation}
Here the $\xi_t$ are independent and identically distributed Gaussian random vectors with
$\E(\xi_t)=0$ and $\E(\xi_t\xi_t^T)=W$.
The deterministic case arises when $W=0$; in this situation the expectation is superfluous,
and we typically consider $\gamma=1$.
In the stochastic case, the covariance $W=W^T>0$ is known and we choose $\gamma\in(0,1)$.

Both stochastic and deterministic LQR problems can be solved using dynamic programming, 
where the value function is first evaluated, and then an optimal controller with linear form is obtained. 
An advantage of the LQR formulation is that the cost-to-go function is a quadratic function of state $x$. 
The value function is obtained by solving a Discrete Algebric Riccati Equation (DARE)
\begin{equation}
	P=Q+\gamma A^TPA-\gamma A^TPB(R+B^TPB)^{-1}B^TPA
\end{equation}
where $P$ induces a positive definite quadratic form which can be interpreted as a value function for the LQR problem. For deterministic LQR, the value function is of the form
\begin{equation}
    V(x_t)=x_t^TPx_t,
\end{equation}
while for stochastic LQR, the value function includes a term independent of the state
\begin{equation}
    V(x_t)=x_t^TPx_t+\frac{\gamma}{1-\gamma}\mathbf{Tr}(WP).
\end{equation}
Once the matrix $P$ has been determined, the unique optimal controller is given by
\begin{equation}
    u^*_t=-Kx_t=-\gamma(R+B^TPB)^{-1}B^TPAx_t.
    \label{controller}
\end{equation}

A companion to the state value function $V$, that takes the state $x$ as input, 
is the state-action value function $Q$, that takes state and action pairs $(x,u)$ as inputs.
It can also be written as a quadratic form. 
For deterministic LQR, the derivation is as follows \citep{adp_lqr}:
\begin{equation}
\begin{aligned}
Q_{K}(x, u) &=r(x, u)+\gamma V_{K} (f(x, u)) \\
&=x^T Q x+u^T R u+\gamma(A x+B u)^T P_{K}(A x+B u) \\
&=[x, u]^T\left[\begin{array}{cc}
Q+\gamma A^T P_{K} A & \gamma A^T P_{K} B \\
\gamma B^T P_{K} A & R+\gamma B^T P_{K} B
\end{array}\right][x, u] \\
&=[x, u]^T\left[\begin{array}{cc}
H_{K(11)} & H_{K(12)} \\
H_{K(21)} & H_{K(22)}
\end{array}\right][x, u] \\
&=[x, u]^T H_{K}[x, u].
\end{aligned}
\label{quadratic_Q}
\end{equation}
Under our hypotheses, the matrix $H_K$ is positive definite.
For stochastic LQR, there is an additional constant term in it
\begin{equation}
Q_{K}(x, u)=[x, u]^{\top} H_{K}[x, u]+\frac{\gamma}{1-\gamma}\mathbf{Tr}(WP_K)
\end{equation}
where $H_K$ has the exact same form as that of deterministic case.

\subsection{Policy iteration for LQR}
Policy iteration is the most important and popular method for model-free control.
Policy iteration involves two repeated steps: policy evaluation, and policy improvement. 
For LQR, both steps have analytical solutions. As shown in (\ref{quadratic_Q}), the cost-to-go function has been parameterized as a quadratic function with respect to state-action pairs. Hence, with the help of the `overbar' function $\overline{[\cdot]}$, the quadratic form can be written as a linear form, and least squares estimation (LSE) can be applied to the policy evaluation step. 

\cite{adp_lqr} first followed this straightforward idea and developed the policy iteration method for deterministic LQR. Given a tuple of data $(x_i,u_i,r_i,x_{i+1})$  belonging to dataset $\mathcal{D}$, 
the following recursion can be written
\begin{equation}
\begin{aligned}
Q_K\left(x_{i}, u_{i}\right)&=r(x_i, u_i)+\gamma V_K(x_{i+1})\\
&=r\left(x_{i}, u_{i}\right)+\gamma Q_K\left(x_{i+1}, K x_{i+1}\right).
\end{aligned}
\end{equation}

With its quadratic representation, the equation for identifying the state-action value function can be constructed
\begin{equation}
\begin{aligned}
\left[x_{i}, u_{i}\right]^{\prime} H_{K}\left[x_{i}, u_{i}\right]&=r\left(x_{i}, u_{i}\right) \\
&-\gamma\left[x_{i+1}, K x_{i+1}\right]^{\prime} H_{K}\left[x_{i+1}, K x_{i+1}\right] 
\label{qbellman}
\end{aligned}
\end{equation}

Using the `overbar' function $\overline{[\cdot]}$ for vectors and matrices, 
equation~(\ref{qbellman}) can be written as
\begin{equation}
    \left(\overline{\left[x_{i}, u_{i}\right]}^{\prime} -\gamma \overline{\left[x_{i+1}, K x_{i+1}\right]}^{\prime}\right) \overline{H}_{K}=r\left(x_{i}, u_{i}\right).
    \label{LSE}
\end{equation}

For the policy improvement step, the updated controller is obtained through
\begin{equation}
    K=-H_{K(22)}^{-1}H_{K(21)}.
\end{equation}

Equation~(\ref{LSE}) is often referred to as \textit{Bellman Residual Minimizing Approximation}. While also utilizing the quadratic structure of the value function, a more advanced approach is proposed in Least Squares Temporal Differences (LSTD) \citep{LSTD} and Least Squares Policy Iteration (LSPI) \citep{LSPI} for the policy evaluation step. The detailed derivation is out of scope for this paper, but the final estimation equation turns out to require only a slight modification of (\ref{LSE}):
\begin{equation}
\begin{aligned}
     \overline{\left[x_{i}, u_{i}\right]}(\overline{\left[x_{i}, u_{i}\right]}^{\prime} - 
     \gamma \overline{\left[x_{i+1}, K x_{i+1}\right]}^{\prime})&\overline{H}_{K}\\
     &\hskip -1em =r\left(x_{i}, u_{i}\right)\overline{\left[x_{i}, u_{i}\right]},
    \end{aligned}
    \label{LSE_advanced}
\end{equation}
where $\overline{\left[x_{i}, u_{i}\right]}(\overline{\left[x_{i}, u_{i}\right]}^{\prime} -\gamma \overline{\left[x_{i+1}, K x_{i+1}\right]}^{\prime})$ is a square matrix. 
Line~(\ref{LSE_advanced}) is often referred as \textit{Least-Squares Fixed-Point Approximation}. 
\citep{LSPI} state that this way of constructing the learning equation requires fewer samples and can obtain a superior policy. As shown above for stochastic LQR, there is a constant term in the value function. In this paper, we still use $\overline{\left[x, u\right]}^{\prime}\overline{H_K}$ to approximate $Q_K$. This won't affect too much on the results under noise with low variance,   which is a reasonable assumption in practice.

In contrast to (\cite{adp_lqr}), LSPI estimates the variable $\overline{H}_{K}$ by summing all equations (\ref{LSE_advanced}) with respect to all the tuples in the dataset $\mathcal{D}$ and then solving the final equation

\begin{equation}
\begin{aligned}
         \sum_{i\in \mathcal{D}}&\overline{\left[x_{i}, u_{i}\right]}(\overline{\left[x_{i}, u_{i}\right]}^{\prime} -\gamma \overline{\left[x_{i+1}, K x_{i+1}\right]}^{\prime}) \overline{H}_{K}\\
         &=\sum_{i\in \mathcal{D}}r\left(x_{i}, u_{i}\right)\overline{\left[x_{i}, u_{i}\right]}.
\end{aligned}
\end{equation}

Furthermore, the framework of LSPI contains an outer loop and an inner loop: 
the inner loop evaluates and improves the policy by iterating with \textit{the same dataset}; 
the outer loop interacts with the environment and updates the dataset with the newest learned policy.

 
\section{Problem Statement }
This paper focuses on the stochastic LQR formulated above, with $W=\sigma_{\omega}^2I$.
Our method can also be adapted to deterministic settings.

We assume that we have, prior to running the algorithm, some information about system dynamics $A$ and $B$. 
We are trying to learn an optimal solution, under the assumption that some of the elements of $A$ and $B$ are  known, and others are unknown.

\section{RL with Partial Knowledge}
In this section, we describe our method of endowing a model-free RL algorithm with partial model knowledge. Our method aims to reduce the sample data consumed by RL, and this is achieved by constructing a better estimation for the value function in the policy evaluation step. Specifically, we utilize optimal control results and elegantly fuse them into the LSPI scheme to enhance learning based LQR. A novel method named Partial Knowledge Least Square Policy Iteration (PLSPI) is developed to consider partial model information and improve sample efficiency in RL. The method is outlined as follows.

\subsection{Representing Partial Model Information}
Our method of considering partial model information is based on decomposing the system dynamics $A$ and $B$ into two parts
\begin{equation}
\begin{aligned}
	A=A_1+A_2\\
	B=B_1+B_2,
\end{aligned}
\end{equation}
where $A_1$ and $B_1$ contain all the known parameters, and $A_2$ and $B_2$ contain all the unknown parameters. Specifically, $A_1$ and $B_1$ set the parameters located in the unknown place as $0$; $A_2$ and $B_2$ set all the known part as 0,
and keeps the unknown part there. However, it should be noted that one can have multiple choices of value to plug-in. If some inaccurate estimation of the unknown parts exits, plug-in the estimated value would be a better choice.

Take a second order scalar system for instance.
Knowing the order of the system, the decomposition can be constructed as 
\begin{equation}
    \begin{array}{c}
         
		x_{n+1}=\left[
	\begin{matrix}
		0& 1\\
		? &?
	\end{matrix}
	\right]x_n+Bu_n
\\
         \downarrow\\
       		x_{n+1}=\left(\left[
	\begin{matrix}
		0& 1\\
		0 &0
	\end{matrix}
	\right]+
	\left[
	\begin{matrix}
		0& 0\\
		? &?
	\end{matrix}
	\right]\right)
	x_n+Bu_n  
    \end{array}
\end{equation}
where the left matrix is $A_1$, the right matrix is $A_2$. It will be illustrated later that without identifying $A_2$, the LQR can still be solved with online data and $A_1$.

With $A_1$ and $B_1$, and a given controller $K_1$, a sub-model containing known model information can be constructed as 
\begin{equation}
    \tilde{x}_1=A_1 x_0 - B_1K_1 x_0
\end{equation}
where $\tilde{x}_1$ represents the consequent state executed from the sub-model, and $x_0$ is the sampled data input to the sub-model. The sub-model shares the same cost matrix $Q$ and $R$ with the original model. The choice of $K_1$ will be discussed in subsection 4.3. 

With the sub-model, a DARE related to $A_1$ and $B_1$ can be obtained as
\begin{equation}
	\begin{aligned}
	P_{K1}=Q+K_1^TRK_1+\gamma(A_1-B_1K_1)^TP_{K1}(A_1-B_1K_1).
	\end{aligned}
\label{subARE1}
\end{equation}
With (\ref{subARE1}), $P_{K1}$ can be solved offline. After having $P_{K1}$, the corresponding state-action value function $H_{K_1}$ can also be calculated offline, as follows:
\begin{equation}
    H_{K_1}=\left[\begin{array}{cc}
R+\gamma A_1^T P_{K_1} A_1 & \gamma A_1^T P_{K_1} B_1 \\
\gamma B_1^T P_{K_1} A_1 & Q+\gamma B_1^T P_{K_1} B_1
\end{array}\right]
\label{H_K1}
\end{equation}
The result will be introduced into reinforcement learning process in the next subsection.

\subsection{Introducing Partial Model Information into RL}
Having $H_{K_1}$, the partial knowledge can be transferred into the RL process. With the property of value function, that the cost-to-go from current state-action can be expanded as one stage cost plus the cost-to-go from the consequent state-action, the following equation can be constructed
\begin{equation}
    (\overline{\left[x_{i}, u_{i}\right]}^{\prime} -\gamma \overline{\left[\tilde{x}_{i+1}, K_1 \tilde{x}_{i+1}\right]}^{\prime}) \overline{H}_{K_1}=r\left(x_{i}, u_{i}\right),
    \label{qbellman_sub}
\end{equation}
where $(x_i,u_i)$ denotes a (state,action) pair sampled from the real system. 
Unlike the state $x_{i+1}$ sampled online, $\tilde{x}_{i+1}$ represents the virtual next state obtained by stepping the sub-model from the current state $x_i$:
\begin{equation}
    \tilde{x}_{i+1}=A_1 x_i - B_1K_1 x_i
    \label{submodel}
\end{equation}

The core of integrating partial model information into RL is achieved by forming the difference between (\ref{qbellman}) and (\ref{qbellman_sub}):
\begin{equation}
\begin{aligned}
    &(\overline{\left[x_{i}, u_{i}\right]}^{\prime} -\gamma \overline{\left[x_{i+1}, K x_{i+1}\right]}^{\prime}) \overline{H}_{K}-(\overline{\left[x_{i}, u_{i}\right]}^{\prime} \\&-\gamma \overline{\left[\tilde{x}_{i+1}, K_1 \tilde{x}_{i+1}\right]}^{\prime}) \overline{H}_{K_1} =r\left(x_{i}, u_{i}\right)-r\left(x_{i}, u_{i}\right)\\
        &\rightrightarrows(\overline{\left[x_{i}, u_{i}\right]}^{\prime} -\gamma \overline{\left[x_{i+1}, K x_{i+1}\right]}^{\prime}) (\overline{H}_{K}-\overline{H}_{K_1})\\&=
        \gamma(\overline{\left[x_{i+1}, K x_{i+1}\right]}^{\prime}- \overline{\left[\tilde{x}_{i+1}, K_1 \tilde{x}_{i+1}\right]}^{\prime}) \overline{H}_{K_1}
\end{aligned}
\end{equation}
Define $\overline{H}_{K}-\overline{H}_{K_1}$ as $\overline{H}_{K_2}$. We will have
\begin{equation}
    \begin{aligned}
    &(\overline{\left[x_{i}, u_{i}\right]}^{\prime} -\gamma \overline{\left[x_{i+1}, K x_{i+1}\right]}^{\prime}) \overline{H}_{K_2}\\&=
        \gamma(\overline{\left[x_{i+1}, K x_{i+1}\right]}^{\prime}- \overline{\left[\tilde{x}_{i+1}, K_1 \tilde{x}_{i+1}\right]}^{\prime}) \overline{H}_{K_1}.
        \label{LSE_partial_simple}
    \end{aligned}
\end{equation}
We hypothesize that this will lead to an improvement on sample efficiency. The idea behind this operation is that instead of identifying $H_K$ from scratch, we can instead identify the gap between what we have known (or a initial guess/baseline of it) and the actual value of it, and hence save data. Note that the right-hand side of the equation can be viewed as a reformulated reward after introducing partial model information. Plus, since the $Q$ function takes state and action as input, $r(x_i,u_i)$ can always be cancelled out no matter what value $K_1$ is fed into the sub-model (\ref{submodel}).

Inspired by LSPI, a more advanced way to construct the LSE equation is used. This can be simply done by slightly adjusting (\ref{LSE_partial_simple})
\begin{equation}
    \begin{aligned}
    &\overline{\left[x_{i}, u_{i}\right]}(\overline{\left[x_{i}, u_{i}\right]}^{\prime} -\gamma \overline{\left[x_{i+1}, K x_{i+1}\right]}^{\prime}) \overline{H}_{K_2}\\&=
        \gamma\overline{\left[x_{i}, u_{i}\right]}(\overline{\left[x_{i+1}, K x_{i+1}\right]}^{\prime}- \overline{\left[\tilde{x}_{i+1}, K_1 \tilde{x}_{i+1}\right]}^{\prime}) \overline{H}_{K_1}.
        \label{LSE_partial_advanced}
    \end{aligned}
\end{equation}

Given a batch of data $\mathcal{D}$, the overall policy evaluation equation is given by
\begin{equation}
    \begin{aligned}
    &\sum_{i \in \mathcal{D}}\overline{\left[x_{i}, u_{i}\right]}(\overline{\left[x_{i}, u_{i}\right]}^{\prime} -\gamma \overline{\left[x_{i+1}, K x_{i+1}\right]}^{\prime}) \overline{H}_{K_2}\\&=
        \gamma\sum_{i \in \mathcal{D}}\overline{\left[x_{i}, u_{i}\right]}(\overline{\left[x_{i+1}, K x_{i+1}\right]}^{\prime}- \overline{\left[\tilde{x}_{i+1}, K_1 \tilde{x}_{i+1}\right]}^{\prime}) \overline{H}_{K_1}.
        \label{LSE_partial_advanced_overall}
    \end{aligned}
\end{equation}
Now (\ref{LSE_partial_advanced_overall}) is the improved policy evaluation equation, that takes both online data and prior model into account. The sample efficiency can benefit from the improved equation.
\subsection{Overall Algorithm}
In (\ref{LSE_partial_advanced}), the choice of $K_1$ is still pending. 
To facilitate a connection between the sub-model and the real model, the preferred choice of $K_1$ is the same controller $K$ that is currently evaluated. However, sometimes $K$ may destabilize the sub-model. When this happens, $K_1$ will be chosen as the optimal controller of the sub-model, which can be obtained by first solving (\ref{subARE1}) and then plugin it into (\ref{controller}). The optimal controller will be noted as $K^*_1$.


      

Another way to choose $K_1$ is to fix it as $K^*_1$ through the whole learning process. We tested this method, and it has similar performance but the hybrid method above works slightly better.

Following LSPI, PLSPI also contains two loops. It is summarized as Algorithm \ref{algo}.

\begin{figure}[h]
		\renewcommand{\algorithmicrequire}{\textbf{Input:}}
		\renewcommand{\algorithmicensure}{\textbf{Output:}}

		\begin{algorithm}[H]
  \begin{algorithmic}[1]
    \State \textbf{Input:} Rollout numbers $N$; Time horizon $T$; Exploration noise $\epsilon_t$; Partial model $A_1$ and $B_1$; Maximum iterations $I_1$ and $I_2$.
      \State \textbf{Initialization:} Initialize controller $K$.
      \For{$i=0,\dots,I_1$}
      \State $\mathcal{D}\leftarrow \emptyset$
      \State Execute $u_t=-Kx_t+\epsilon_t$ for $N$ episodes. Store $N\times T$ pairs of data for $\mathcal{D}$.
      \For{$j=0,\dots,I_2$}
        \If {$A_1-B_1K$ is stable}
        \State $K_1=K$
        \Else 
        \State $K_1=K^*_1$
        \EndIf
        \State Solve (\ref{subARE1})(\ref{H_K1}) to obtain $H_{K1}$.
        \State Identify $H_{K_2}$ with (\ref{LSE_partial_advanced_overall}), and obtain $H_K$ with $H_K=H_{K_1}+H_{K_2}$.
        \State Improve policy with $K=-H_{K(22)}^{-1}H_{K(21)}$.
      \EndFor
      \EndFor
      \State \textbf{return} $K$
  \end{algorithmic}
			\caption{Partial Knowledge LSPI (PLSPI)}
   \label{algo}
		\end{algorithm}
	\end{figure}
\begin{remark}
Our method can serve as a bridge between control and RL.
    Consider the extreme cases. When the prior information is the full model, and the initial controller is initialized with the full model, the estimation actually needs no data: it reduces to solving a fully-specified optimal control problem. When prior information is completely absent, then the scheme reduces  to classic LSPI, which finds the optimal policy totally by data.
\end{remark}


\section{Numerical Examples}
We present several simulation examples to illustrate our
algorithm. The first example tests the effectiveness of LSPI and the proposed method on LQR, specifically on a deterministic undiscounted setting. The second example is a more complex and unstable model with practical origins (cooling system) which is used for comparison purposes. In all experiments, the exploration noise follows \emph{i.i.d.} Gaussian random vectors with
$\E(\epsilon_t)=0$ and $\E(\epsilon_t\epsilon_t^T)=\sigma_{\eta}^2I$, with $\sigma_{\eta}^2=0.1$. The controller $K$ is initialized as $0$.

\subsection{Example 1}

Consider the classic discrete-time double integrator with dynamic coefficients
\begin{equation*}
A=
\left[
 \begin{matrix}
    1&1\\
    0&1
\end{matrix}\right],
\quad
B=
\left[
 \begin{matrix}
    0\\
    1
\end{matrix}\right],
\end{equation*}
and quadratic cost coefficients
\begin{equation*}
Q=
\left[
 \begin{matrix}
    1&0\\
    0&1
\end{matrix}\right],
\quad
R=1.
\end{equation*}
The system is deterministic so that the noise $\xi_t=0$. The discount factor $\gamma$ is set to $1$. 
The eigenvalues of the system both equal $1$, meaning that the system is on the boundary of stability.

For our PLSPI method, the partial information is set as
\begin{equation*}
A_1=
\left[
 \begin{matrix}
    1&1\\
    0&0
\end{matrix}\right],
\qquad
B_1=
\left[
 \begin{matrix}
    0\\
    1
\end{matrix}\right].
\end{equation*}

The problem is simple, so both LSPI and PLSPI perform well. 
They converge to the optimal controller within just 1 or 2 iterations. 
Each iteration contains 30 rollouts, with one rollout being a run with a horizon of 20 steps. 

\subsection{Example 2}
Next we consider a simplified model of a 3-level cooling system, which is a popular test scenario for RL applications \citep{LSTD_lqr,LSPI_lqr,model_based_lqr}. The detail is as follows,
\begin{equation*}
A=
\left[
 \begin{matrix}
    1.01&0.01&0\\
    0.01&1.01&0.01\\
    0&0.01&1.01
\end{matrix}\right],   
\end{equation*}
$B=I$, $Q=I$, $R=1000I$, $\sigma_{\omega}^2=0.01$, and $\gamma=0.98$.
The open-loop system is unstable, which increases the difficulty for the learning algorithm. 
To make the problem harder, we use the scale factor $1000$ to penalize control action through $R$ much more heavily than state offsets through $Q$ 
This puts the maximum eigenvalue of the optimal closed-loop around $0.98$, making the system challenging to tune. 
We assume knowledge of only the diagonal entries in the dynamic matrices,
setting the partial information as
    \begin{equation*}
    A_1=
    \left[
     \begin{matrix}
        1.01&0&0\\
        0&1.01&0\\
        0&0&1.01
    \end{matrix}\right],
    \qquad
    B_1=
    \left[
     \begin{matrix}
        1&0&0\\
        0&1&0\\
        0&0&1
    \end{matrix}\right].  
    \end{equation*}
    
\begin{figure}[htbp]
    \centering
    \includegraphics[width=3.2in]{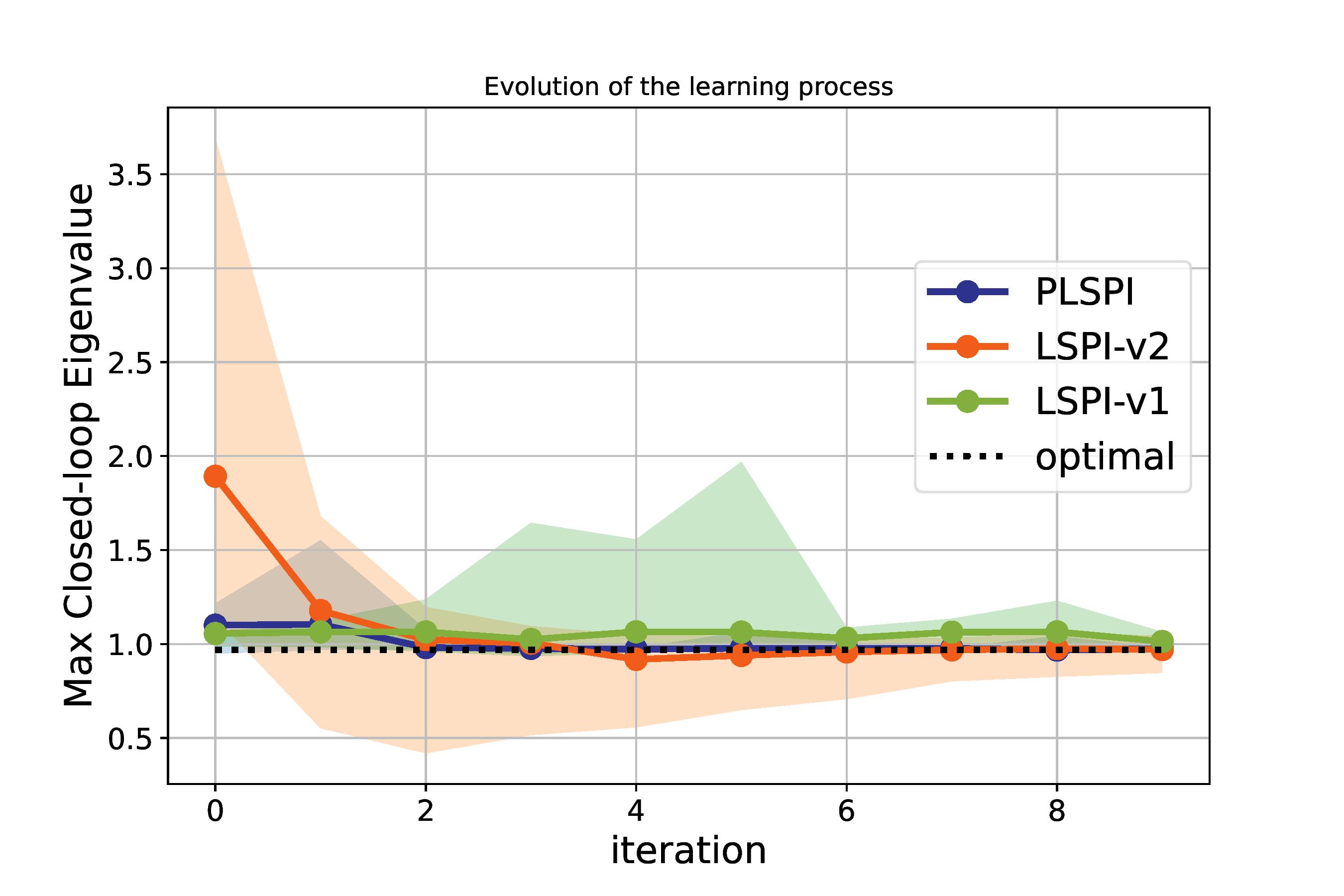}
    \caption{The comparison between different methods in terms of the evolution of closed-loop system with respect to iterations. Each iteration contains 30 rollouts, with each rollout running 20 steps. The shaded area represents the variance of each method, with percentile 0-75\%. The solid line represents the median of each method. }
    \label{eigen}
\end{figure}

The performance of PLSPI and LSPI is compared in terms of the average trend and the variation of the evolution from multiple simulations. Specifically, we run LSPI and our PLSPI for 10 times each, and obtain Fig.~\ref{eigen}.

Particularly, two versions of LSPI are tested here. LSPI-v1 follows the original setting, where the inner loop iterates for multiple times (5 in our case) given a fixed batch of data; LSPI-v2 adopts the setting in \citep{LSPI_lqr}, where there is no inner loop, meaning each batch of data is only used for once.

Fig.~\ref{eigen} shows PLSPI converging faster than both versions of LSPI. More importantly, the shaded region showing observed variance is also smaller for PLSPI, which indicates another benefit: given limited data, PLSPI learns the optimal controller with better accuracy. These benefits contribute to the better sample efficiency of our method.

\begin{figure}[htbp]
    \centering
    \includegraphics[width=3.2in]{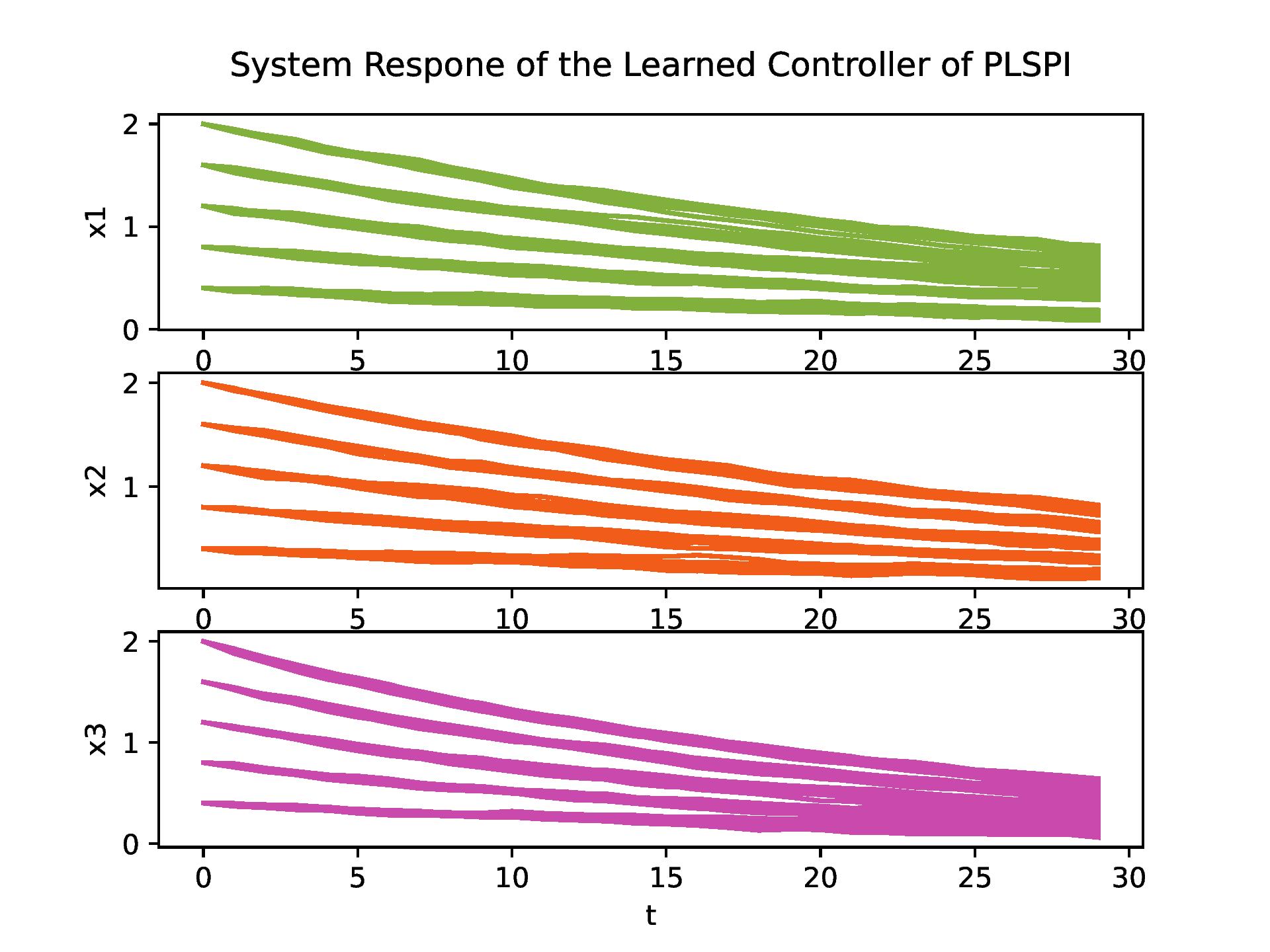}
    \caption{Stochastic trajectories generated from the learned controller. The $X$-axis represents simulation steps.}
    \label{response2}
\end{figure}

The stochastic system response is also tested, with the results shown in Fig.\ref{response2}. Trajectories in Fig.\ref{response2} show that the learned controller can regulate the states under a stochastic system point of view.

\begin{figure}[htbp]
    \centering
    \includegraphics[width=3.2in]{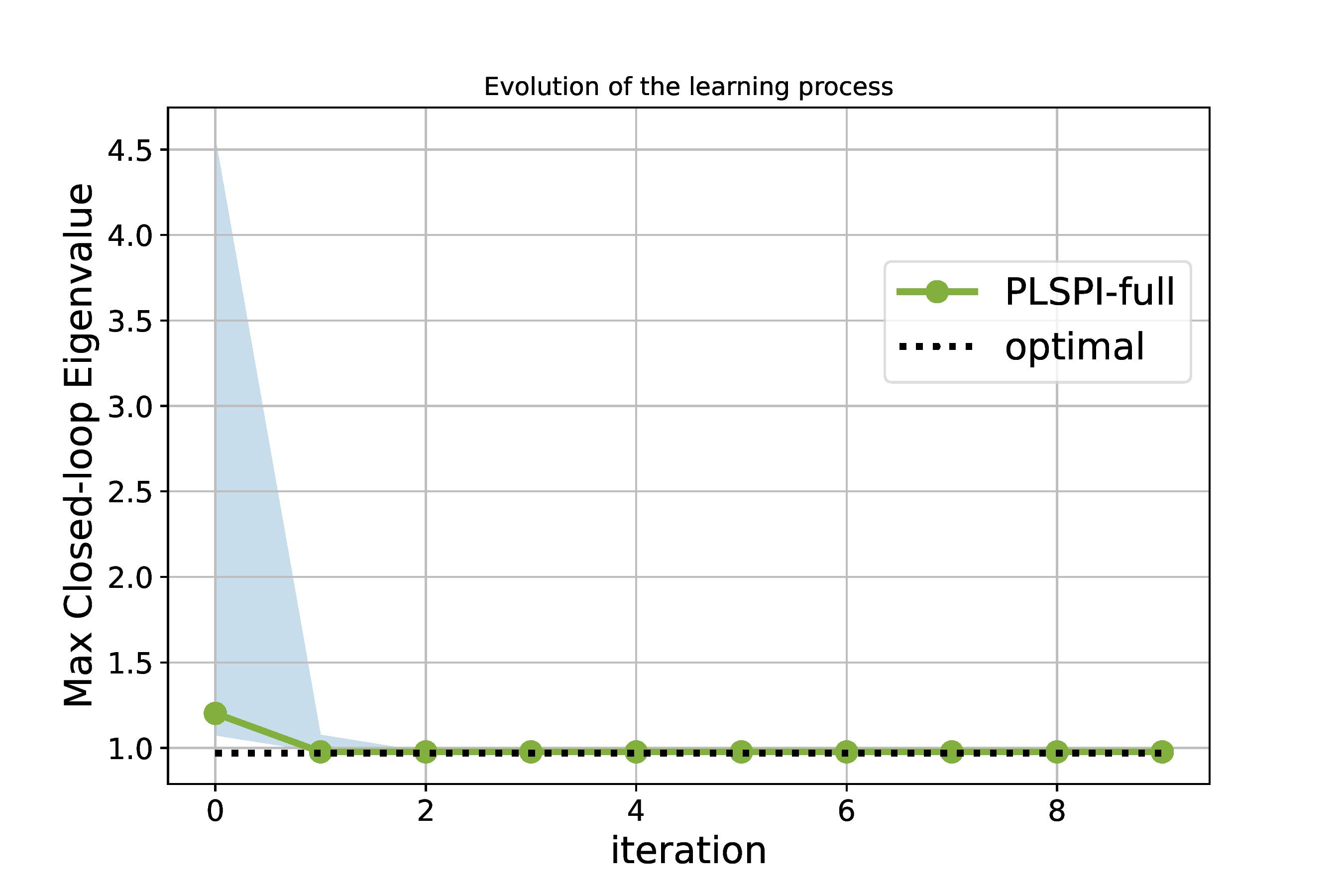}
    \caption{Results of PLSPI given a completely known system model, with percentile 0-100\%.}
    \label{eigenfull}
\end{figure}

We further apply the algorithm to a completely known system model (knowledge on noise term is not required) to test an extreme case. Fig.\ref{eigenfull} shows the result.

In Fig. \ref{eigenfull}, the learner gets to the optimal solution with only 1 iteration, and evolves without any variance. The initial variance is related to the random initialization of the controller. If the controller is initialized with the given model, the variance can also be eliminated. This result follows the purpose of the algorithm design: given complete information about the system, the algorithm will follow a fully optimal control solution process, without consuming any data.
\section{Conclusion}
In control tasks, some partial parametric model information is often known, but seemingly under-utilized in model-free RL. Taking it as a starting point, we have adopted complementary benefits  in model-free RL and model-based control to develop a framework that aims to bridge both sides. This paper is a proof of concept and there are many avenues to explore. These include the exact learning for value function under stochastic setting; theoretical guarantee on when partial information is useful; and extensions to nonlinear systems. We believe this is a promising area to investigate further as RL gains traction in process systems engineering.

\bibliography{ifacconf}             
                                                   







\end{document}